\documentclass[11pt,twoside]{article}

\usepackage{asp2014}

\aspSuppressVolSlug
\resetcounters

\begin{document}

\title{TAP and the Data Models}

\author{L. Michel,$^1$ F. Bonnarel,$^2$ M. Louys, $^{2,3}$ and D. Morris$^4$}
\affil{$^1$Universit\'e de Strasbourg, Observatoire Astronomique de Strasbourg, CNRS UMR 7550, Strasbourg, France; \email{laurent.michel@astro.unistra.fr}}
\affil{$^2$Universit\'e de Strasbourg, CDS/Observatoire Astronomique de Strasbourg, CNRS UMR 7550, Strasbourg, France}
\affil{$^3$Universit\'e de Strasbourg, ICube, CNRS UMR 7357, Strasbourg, France }
\affil{$^4$University of Edinburgh, Edinburgh, Scotland, United Kingdom} 

\paperauthor{Laurent Michel}{laurent.michel@astro.unistra.fr}{0000-0001-5702-0019}{University of Strasbourg}{Observatoire Astronomique, UMR7550-CNRS}{Strasbourg}{}{67200}{France}
\paperauthor{Mireille Louys}{mireille.louys@unistra.fr}{0000-0002-4334-1142}{University of Strasbourg}{ICube UMR7357-CNRS, Observatoire Astronomique UMR7357-CNRS} {Strasbourg}{}{67000}{France}
\paperauthor{Fran\c{c}ois Bonnarel}{francois.bonnarel@astro.unistra.fr}{}{University of Strasbourg}{Observatoire Astronomique, UMR7550-CNRS}{Strasbourg}{}{67000}{France} 
\paperauthor{D. Morris}{dmr@roe.ac.uk}{}{Institute for Astronomy}{University of Edinburgh}{Edinburgh}{Scotland}{EH93HJ}{United Kingdom} 



\begin{abstract}
The purpose of the "TAP and the Data Models" Bird of Feathers session was to discuss the relevance of enabling TAP services to deal with IVOA standardized data models and to refine the functionalities required to implement such a capability.
\end{abstract}



\section{Introduction}
TAP \citep{2019ivoa.spec.0927D} is one of the big achievements of the VO. This protocol gives any relational database a high level of interoperability thanks to 3 IVOA standards: 1) the TAP\_SCHEMA that describes tables and the way to join them, 2) ADQL \citep{2008ivoa.spec.1030O}, a subset of SQL, with some astronomy-specific features and 3) UWS \citep{2016ivoa.spec.1024H}, a  REST API to handle service requests.

These features provide a common way to discover the content of TAP services and to query them.
This works very well with relational data and we propose to investigate the possibility for TAP services to map searched data on data models. Indeed several data models have been developed by IVOA in order to tackle the complexity of the relationships between astronomical data features. Among those we can quote Photometry Data Model \citep{2013ivoa.spec.1005S}, Coordinates \citep{2021ivoa.spec.Coords}, Measurements \citep{2021ivoa.spec.Meas} or MANGO \citep{github:mango} that is  well suited to describe astronomical source properties and relations to some data sets representing these sources. TAP services are able to host complex data bound with joins but the standard still misses important features to serve real model instances: 1) Clients must be able to discover whether model views are available for a given resource 2) TAP services must be able to  host extra meta-data necessary to build model instances on the fly 3) TAP services must support serialization formats suitable for complex data

The purpose of the BoF \footnote{https://www.youtube.com/watch?v=HSWTgv7blfM} is to discuss the relevance of enabling TAP services to deal with Data Models and to sketch up the functionalities required to implement such a capability. Possible strategies are described elsewhere in this conference proceedings   \citep{X4-010_adassxxxi} 

\section{Browsing data built upon any relational schema}
Some TAP services are already able to serve data that are built upon complex relational schemes.
We can mention TAP Simbad \citep{2015ASPC..495..429O} built upon an internal relational schema or all services based on CAOM \citep{2019ASPC..521..446D} which is a model published out of the VO standard process. Another popular case is the relational registry  based on a relational schema published as a VO standard \citep{2014ivoa.spec.1208D}

In the case of services based on relational schemata, 
table data are connected together by joins that are discoverable in the TAP\_SCHEMA.

Although there is no standard way to link such schemes to a model vision as VODML \citep{2018ivoa.spec.0910L} can provide, the services mentioned above are relevant use-cases to prospect different methods to deal with modeled data in a TAP context.

This capability is being exercised in the frame of the TAP-Complex project 
\footnote{https://github.com/lmichel/TAP-complex-data}
which aims at providing TAPHandle  \citep{2014ASPC..485...15M} with some advanced model ability. This Javascript proof of concept is based on a middle-ware API that explores the TAP\_SCHEMA to map all joins related to the searched table. The rows of that table are displayed in a usual way, but when the user clicks on one of them, the names of the joined tables are listed below and joined data can be fetched by clicking on those table labels. The mapping of the table joins done at connection time makes the access to joined data very easy. Users can set constraints on any table and query strings are automatically generated with all relevant join statements. Query results are re-normalized internally. 

\section{Legacy data annotation}

A step toward a better DM integration in TAP consists in enabling services to annotate legacy data either by data recombination and grouping or by providing complete model views. 
 This requires the server to operate a post-processing inserting into VOTables annotations that bind data columns with model leaves.

This can be done by using GROUPs and UTypes as shown by J.Silverman (Caltech). The Caltech/IPAC-IRSA TAP service caches information  about interesting combinations of database-table columns at start-up.  At run time, the service adds a GROUP element  to VOTable-format responses for any such group of columns all of which are contained in the expanded SELECT clause of the user query. 
Judith's presentation showed how this functionality is supported. An open question is to know how users can discover that e.g. the exposed source table is linked with a detection table. Without a datamodel mapping, 
TAP provides no way to say "service is exposing sources with their different detections".

The IVOA is working on a more generic solution based on a mapping syntax \citep{github:mapping} that allows to map data on any model compliant with the VODML meta-model. These annotations are built as leading XML  blocks in VOTables. Such blocks denote the model structure and contains references to the appropriate table FIELDs. Model-aware clients can build model instances just by reading the annotation block and by resolving the FIELD references to get the model leaf values. In a TAP context, the server must be able to automatically generate such annotations. For this, it must check that the selected columns match with the model definitions and thus can be mapped on that model. To operate the mapping, the server needs further information such as coordinate frames and data profile resources giving the binding between table columns and model leaves. A prototype 
\citep{X3-010_adassxxxi} implementing this feature has been demonstrated \footnote{https://github.com/lmichel/TAP-annoter}. 

\section{Object Relational Mapping (ORM) Strategy}

TAP services can also be used to host model instances. In this case, we must not map data on a model anymore but we have to do a real object relational mapping. However, proposing a common ORM schema is not on the VO roadmap. The work around strategy is to propose one specific standard per model. This has been done first for ObsTAP \citep{2011ivoa.spec.1028T} which flattens the ObsCore model on one table. This is also the case for ProvTAP \footnote{https://wiki.ivoa.net/internal/IVOA/ObservationProvenanceDataModel/ProvTAP.pdf} which  proposes a relational view for Provenance \citep{2020ivoa.spec.0411S} data. A prototype (ProvHiPS) tracing the provenance of HST HiPS tiles has been demonstrated. As the model mapping is defined by a standard, there is no need to add extra information to the TAP service. Both TAP\_SCHEMA content and  meta-data defined in that standard provide all pieces of information needed to construct model instances from query results. There are however 2 major issues: 1) Provenance instances cannot be serialized in one single table; in order to solve this issue resulting VOTable documents must  either contain multiple tables or provide a flattened view of the model itself (namely last step provenance)
2) The client must be able to tell the server it is searching Provenance instances.    

\section{Conclusions and Perspective} 
This session and the following discussions \footnote{https://wiki.ivoa.net/internal/IVOA/TapandTheDMs/Etherpad\_notes.txt} highlighted that TAP services can already serve hierarchical data and that serving legacy data with annotations or even Provenance instances is within our reach. On server side, we need to add model profiles and other extra meta-data to the TAP\_SCHEMA to allow the addition of model annotations into the query responses. This post processing could also add other tables with associated data  (e.g. sources with their detections). Such a feature wouldn't break existing TAP clients. The ADQL queries would remain unchanged and legacy clients could simply skip annotations.

\acknowledgements 
We would like to thank Judith Silverman (Caltech) for her contribution as well as the large audience for  participating and for the interesting discussions that have followed the session.

\bibliography{B3-001} 
\end{document}